\documentclass[modern, dvipsnames]{aastex631}
\usepackage{CJK}
\usepackage{microtype}
\usepackage{savesym}
  \savesymbol{lambdabar}
\usepackage{newpxtext, eulerpx}
  \restoresymbol{newpx}{lambdabar}
\usepackage[T1]{fontenc}
\usepackage{fontawesome}
\usepackage{amsmath}
\usepackage{bm}
\usepackage{nicefrac}
\usepackage[caption=false]{subfig}
\usepackage[figure,figure*]{hypcap}
\usepackage{tikz}
  \usetikzlibrary{positioning, fit, calc, arrows.meta}
\usepackage{booktabs}

\newcommand{\pmwd}{{\usefont{T1}{nova}{m}{sl}pmwd}}

\newcommand{\gkai}[1]{\begin{CJK*}{UTF8}{gkai}\raisebox{.1em}{(}#1\raisebox{.1em}{)}\end{CJK*}}

\newcommand{\deltaD}{\delta^\textsc{d}}
\newcommand{\deltaK}{\delta^\textsc{k}}
\renewcommand{\d}{d}
\newcommand{\p}{\partial}
\newcommand{\cJ}{\mathcal{J}}
\newcommand{\cR}{\mathcal{R}}
\newcommand{\cL}{\mathcal{L}}

\bmdefine{\vzero}{0}
\bmdefine{\vI}{I}
\bmdefine{\vnabla}{\nabla}
\bmdefine{\vtheta}{\theta}  % parameters
\bmdefine{\vomega}{\omega}  % white noise modes
\bmdefine{\vk}{k}  % wavevectors
\bmdefine{\vx}{x}  % comoving and canonical coordinates
\bmdefine{\vq}{q}  % Lagrangian coordinates
\bmdefine{\vs}{s}  % displacements
\bmdefine{\vp}{p}  % canonical momenta
\bmdefine{\va}{a}  % accelerations
\bmdefine{\vz}{z}  % states
\bmdefine{\vf}{f}
\bmdefine{\vF}{F}
\bmdefine{\vDelta}{\Delta}
\bmdefine{\vLambda}{\Lambda}
\bmdefine{\vlambda}{\lambda}
\bmdefine{\vvarphi}{\varphi}
\bmdefine{\vxi}{\xi}  % x adjoint
\bmdefine{\vpi}{\pi}  % p adjoint
\bmdefine{\valpha}{\alpha}  % force vjp x gradient
\bmdefine{\vzeta}{\zeta}  % force vjp theta gradient
\newcommand{\half}{\nicefrac12}
\newcommand{\As}{A_\mathrm{s}}
\newcommand{\ns}{n_\mathrm{s}}
\newcommand{\Omegam}{\Omega_\mathrm{m}}
\newcommand{\Omegab}{\Omega_\mathrm{b}}
\newcommand{\Mpc}{\mathrm{Mpc}}
\newcommand{\ic}{\mathrm{i}}
\newcommand{\Plin}{P_\mathrm{lin}}

\newcommand{\GPU}{NVIDIA H100 PCIe}

\begin{document}

\title{\large Differentiable Cosmological Simulation with the Adjoint Method
\vspace{0.3em}}

\author[0000-0002-0701-1410]{\normalsize Yin Li \gkai{李寅}}
\affiliation{Department of Mathematics and Theory, Peng Cheng
Laboratory, Shenzhen, Guangdong 518066, China}
\affiliation{Center for Computational Mathematics, Flatiron Institute,
New York, New York 10010, USA}
\affiliation{Center for Computational Astrophysics, Flatiron Institute,
New York, New York 10010, USA}

\author[0000-0002-1670-2248]{\normalsize Chirag Modi}
\affiliation{Center for Computational Mathematics, Flatiron Institute,
New York, New York 10010, USA}
\affiliation{Center for Computational Astrophysics, Flatiron Institute,
New York, New York 10010, USA}

\author[0000-0001-5044-7204]{\normalsize Drew Jamieson}
\affiliation{Max Planck Institute for Astrophysics, 85748 Garching bei
M\"unchen, Germany}

\author[0000-0002-9300-2632]{\normalsize Yucheng Zhang \gkai{张宇澄}}
\affiliation{Department of Mathematics and Theory, Peng Cheng
Laboratory, Shenzhen, Guangdong 518066, China}
\affiliation{Center for Cosmology and Particle Physics, Department of
Physics, New York University, New York, New York 10003, USA}

\author[0000-0003-0745-9431]{\normalsize Libin Lu \gkai{陆利彬}}
\affiliation{Center for Computational Mathematics, Flatiron Institute,
New York, New York 10010, USA}

\author[0000-0001-5590-0581]{\normalsize Yu Feng \gkai{冯雨}}
\affiliation{Berkeley Center for Cosmological Physics, University of
California, Berkeley, California 94720, USA}

\author[0000-0001-7956-0542]{\normalsize Fran\c{c}ois Lanusse}
\affiliation{AIM, CEA, CNRS, Universit\'e Paris-Saclay, Universit\'e
Paris Diderot,\\ Sorbonne Paris Cit\'e, F-91191 Gif-sur-Yvette, France}

\author[0000-0003-2895-8715]{\normalsize Leslie Greengard}
\affiliation{Center for Computational Mathematics, Flatiron Institute,
New York, New York 10010, USA}
\affiliation{Courant Institute, New York University, New York, New York
10012, USA}

\shorttitle{Differentiable Simulation with Adjoint Method}
\shortauthors{Li et al.}

\correspondingauthor{Yin Li}
\email{eelregit@gmail.com}

\begin{abstract}

Rapid advances in deep learning have brought not only myriad powerful
neural networks, but also breakthroughs that benefit established
scientific research.
In particular, automatic differentiation (AD) tools and computational
accelerators like GPUs have facilitated forward modeling of the Universe
with differentiable simulations.
Based on analytic or automatic backpropagation, current differentiable
cosmological simulations are limited by memory, and thus are subject to
a trade-off between time and space/mass resolution, usually sacrificing
both.
We present a new approach free of such constraints, using the adjoint
method and reverse time integration.
It enables larger and more accurate forward modeling at the field level,
and will improve gradient based optimization and inference.
We implement it in an open-source particle-mesh (PM) $N$-body library
\pmwd\ (particle-mesh with derivatives).
Based on the powerful AD system \texttt{JAX}, \pmwd\ is fully
differentiable, and is highly performant on GPUs.

\clearpage
\end{abstract}

\section{Introduction}

Current established workflows of statistical inference from cosmological
datasets involve reducing cleaned data to summary statistics like the
power spectrum, and predicting these statistics using perturbation
theories, semi-analytic models, or simulation-calibrated emulators.
These can be suboptimal due to the limited model fidelity and the
risk of information loss in data compression.
Cosmological simulations \citep{HockneyEastwood1988, AnguloHahn2022} can
accurately predict structure formation even in the nonlinear regime at
the level of the fields.
Using simulations as forward models also naturally accounts for the
cross-correlation of different observables, and can easily incorporate
systematic errors.
This approach has been intractable due to the large computational costs
on conventional CPU clusters, but rapid advances in accelerator
technology like GPUs open the possibility of simulation-based modeling
and inference \citep{CranmerEtAl2020}.
Furthermore, model differentiability enabled by AD libraries can
accelerate parameter constraint with gradient-based optimization and
inference.
A differentiable field-level forward model combining these two features
is able to constrain physical parameters together with the initial
conditions of the Universe.

The first differentiable cosmological simulations, such as BORG, ELUCID,
and BORG-PM \citep{BORG, ELUCID, BORG-PM}, were developed before the
advent of modern AD systems, and were based on the analytic derivatives,
which involve first convoluted derivation by hand using the chain rule
\citep[see e.g.,][App.~D]{SeljakEtAl2017} before implementing them in
code.
Later codes including \texttt{FastPM} and \texttt{FlowPM} \citep{FastPM,
vmad, SeljakEtAl2017, FlowPM} compute gradients using the AD engines,
namely \texttt{vmad} (written by the same authors) and
\texttt{TensorFlow}, respectively.
The AD frameworks automatically apply the chain rule to the primitive
operations that comprise the whole simulation, relieving the burden of
derivation and implementation of the derivatives.
Both analytic differentiation and AD backpropagate the gradients through
the whole history, which requires saving the states at all time steps in
memory.
Therefore, they are subject to a trade-off between time and space/mass
resolution, usually sacrificing both.
As a result, they lose accuracy on small scales and in dense regions
where the time resolution is important, e.g., in weak lensing
\citep{MADLens}.

\begin{figure*}[t]
\centering
\tikzstyle{node style}=[
  node distance=2em, minimum size=2em, text centered,
  to/.style={-to, gray, thick},
  toto/.style={double distance=1pt, -Implies, gray, thick},
  from/.style={to-, gray, thick},
]
\tikzstyle{box style}=[
  rectangle, inner sep=0.4em, rounded corners, draw=gray, thick,
]
\subfloat[model structure]{
\label{fig:model}
\begin{tikzpicture}[node style]
  \node (boltz) {Boltz};
  \node (ic) [right=of boltz] {LPT} edge [from] (boltz);
  \node (integ) [right=of ic] {Integ} edge [from] (ic);
  \node (obs) [below=of integ] {Obs}
        edge [from] (ic)
        edge [from] (integ);
  \node (obj) [below=of obs] {Obj} edge [from] (obs);
  \node (param) [below=of ic] {$\vtheta, \vomega$}
        edge [to] (boltz)
        edge [toto] (ic)
        edge [to] (integ)
        edge [to] (obs)
        edge [toto] (obj);
  \node (box) [box style, fit=(integ) (obs)] {};
  \useasboundingbox ($ (box.north) + (0, 1em) $);
\end{tikzpicture}
}
\hfill
\subfloat[integration-observation loop]{
\label{fig:loop}
\begin{tikzpicture}[node style]
  \node (ic) {LPT};
  \node (integ_1) [right=of ic] {$I_1$} edge [from] (ic);
  \node (obs_1) [below=of integ_1] {$O_1$}
        edge [from] (ic)
        edge [from] (integ_1);
  \node (integ_1i) [right=of integ_1] {$\cdots$} edge [from] (integ_1);
  \node (obs_1i) [below=of integ_1i] {$\cdots$}
        edge [from] (integ_1i)
        edge [from] (obs_1);
  \node (integ_i) [right=of integ_1i] {$I_i$} edge [from] (integ_1i);
  \node (obs_i) [below=of integ_i] {$O_i$}
        edge [from] (integ_i)
        edge [from] (obs_1i);
  \node (integ_in) [right=of integ_i] {$\cdots$} edge [from] (integ_i);
  \node (obs_in) [below=of integ_in] {$\cdots$}
        edge [from] (integ_in)
        edge [from] (obs_i);
  \node (integ_n) [right=of integ_in] {$I_n$} edge [from] (integ_in);
  \node (obs_n) [below=of integ_n] {$O_n$}
        edge [from] (integ_n)
        edge [from] (obs_in);
  \node (obj) [below=of obs_n] {Obj} edge [from] (obs_n);
  \node (box) [box style, fit=(integ_1) (obs_n)] {};
  \node [box style, inner sep=0em, dashed, fit=(integ_i) (obs_i)] {};
  \useasboundingbox ($ (box.north) + (0, 1em) $);
\end{tikzpicture}
}
\\
\hfill
\subfloat[$i$-th (leapfrog) integration-observation step]{
\label{fig:step}
\begin{tikzpicture}[node style, node distance=2em and 1.5em]
  \node (kick_a) {$K_{i-1}^{i-\half}$};
  \node (drift) [right=of kick_a] {$D_{i-1}^i$} edge [from] (kick_a);
  \node (force) [right=of drift] {$F_i$} edge [from] (drift);
  \node (kick_b) [right=of force] {$K_{i-\half}^i$}
        edge [from] (force);
  \node (obs) [below=of kick_b] {$O_{i-1}^i$} edge [from] (kick_b);
  \node (integ_im1) [left=of kick_a, minimum width=0em] {}
        edge [to] (kick_a);
  \node (integ_ip1) [right=of kick_b, minimum width=0em] {}
        edge [from] (kick_b);
  \node (obs_im1) [below=of integ_im1, minimum width=0em] {}
        edge [to] (obs);
  \node (obs_ip1) [below=of integ_ip1, minimum width=0em] {}
        edge [from] (obs);
  \node (box) [box style, dashed, fit=(kick_a) (kick_b) (obs)] {};
  \useasboundingbox ($ (box.south) - (0, 1em) $) --
                    ($ (box.north) + (0, 1em) $);
\end{tikzpicture}
}
\caption{Simulation-based forward model of the Universe.
\subref{fig:model} shows the overall model structure.
Single arrows from the cosmological parameters $\vtheta$ and white noise
modes $\vomega$ indicate dependence on $\vtheta$ only, while double
arrows imply dependence on both.
The time integration loop in \subref{fig:loop} expands the solid box in
\subref{fig:model}, and a single time step in \subref{fig:step} further
expands the dashed box in \subref{fig:loop}.
We describe different operators in \autoref{sec:fwd}: Boltzmann solver
(``Boltz'') and initial condition generator by the Lagrangian
perturbation theory (``LPT'') in \autoref{sec:ic}; force solver ($F$) in
\autoref{sec:force}; time integration (``Integ''), kick ($K$), and drift
($D$) in \autoref{sec:integ}; observation (``Obs'' and $O$) and
objective (``Obj'') in \autoref{sec:obsobj}.
Gradients flow backward with all arrows reversed (\autoref{sec:adj}).
}
\end{figure*}
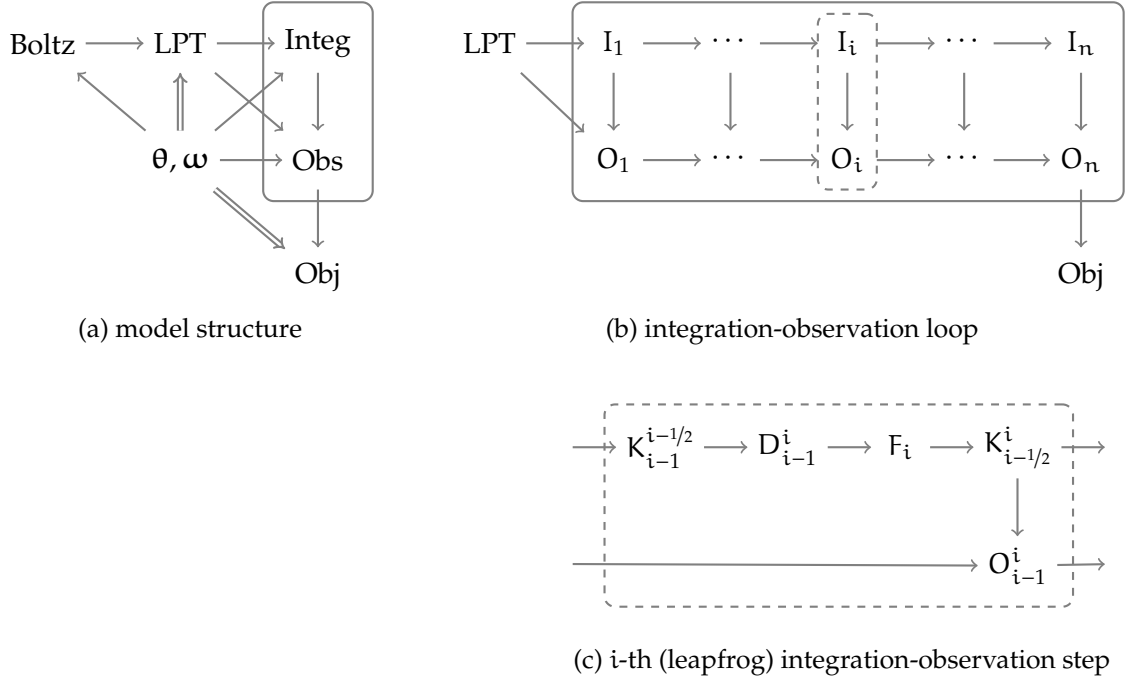

Alternatively, the adjoint method provides systematic ways of deriving
the gradients of an objective function $\cJ$ under constraints
\citep{Pontryagin1962}, such as those imposed by the $N$-body equations
of motion in a simulated Universe.
It identifies a set of \emph{adjoint variables} $\vlambda$, dual to the
state variables $\vz$ of the model, and carrying the gradient
information of the objective function with respect to the model state
$\p\cJ / \p\vz$.
For time-dependent problems, the adjoint variables evolve
\emph{backward} in time by a set of equations dual to that of the
forward evolution, known as the \emph{adjoint equations}.
For continuous time, the adjoint equations are a set of differential
equations, while in the discrete case they become difference equations
which are practically a systematic way to structure the chain rule or
backpropagation.
Their initial conditions are set by the explicit dependence of the
objective on the simulation state, e.g., $\vlambda_n = \p\cJ / \p\vz_n$
if $\cJ$ is a function of the final state $\vz_n$.
Solving the adjoint equations can help us to propagate the gradient
information via the adjoint variables to the input parameters $\vtheta$,
to compute the objective gradients $\d\cJ / \d\vtheta$.
And we will see later that the propagated and accumulated gradients on
parameters come naturally from multiple origins, each reflecting a
$\vtheta$-dependence at one stage of the modeling in
\autoref{fig:model}.

The backward adjoint evolutions depend on the states in the forward run,
which we can re-simulate with reverse time integration if the dynamics
is reversible, thereby dramatically reducing the memory cost
\citep{NeuralODE}.
Furthermore, we derive the discrete adjoint equations dual to the
discrete forward time integration, known as the discretize-then-optimize
approach \citep[e.g.,][]{ANODE}, to ensure gradients propagate backward
along the \emph{same} trajectory as taken by the forward time
integration.
This is in contrast with the optimize-then-discretize approach, that
numerically integrates the continuous adjoint equations, and is prone to
larger deviations between the forward and the backward trajectories due
to different discretizations \citep{LanzieriLanusseEtAl2022}.
In brief, to compute the gradient we only need to evolve a simulation
forward, and then backward jointly with its dual adjoint equations.
We introduce the adjoint method first for generic time-dependent
problems in both continuous and discrete cases in \autoref{sec:bwd}, and
then present its application on cosmological simulation in
\autoref{sec:adj}.

We implement the adjoint method with reverse time integration in a new
differentiable PM library \pmwd\ using \texttt{JAX} \citep{pmwd}.
\pmwd\ is memory efficient at gradient computation with a space
complexity independent of the number of time steps, and is computation
efficient when running on GPUs.

\vspace{1em}
\section{Forward Simulation}
\label{sec:fwd}

We first review and formulate all components of the $N$-body
simulation-based forward model of the cosmological structure formation.

\vspace{1em}
\subsection{Initial Conditions \& Perturbation Theories}
\label{sec:ic}

$N$-body particles discretize the uniform distribution of matter at the
beginning of cosmic history (scale factor $a(t) \to 0$) at their
Lagrangian positions $\vq$, typically on a Cartesian grid, from which
they then evolve by displacements $\vs$ to their later positions, i.e.,
\begin{equation}
\vx = \vq + \vs(\vq).
\end{equation}
To account for the cosmic background that expands globally, while $\vx$
is the comoving position relative to this background, the physical
position grows with the expansion scale factor, i.e., $a \vx$.
The expansion rate is described by the Hubble parameter $H \triangleq \d
\ln a / \d t$.

The initial conditions of particles can be set perturbatively when
$|\vnabla \cdot \vs|$, the linear approximation of the density
fluctuation, is much less than 1.
We compute the initial displacements and momenta using the second order
Lagrangian perturbation theory \citep[2LPT,][]{BouchetEtAl1995}:
\begin{align}
\vs &= D_1 \vs^{(1)} + D_2 \vs^{(2)}, \nonumber\\
\vp &= a^2 \dot\vx = a^2 \dot\vs
  = a^2 H \bigl( D_1' \vs^{(1)} + D_2' \vs^{(2)} \bigr),
\label{lpt}
\end{align}
where $\vp$ is the canonical momentum\footnote{We omit the particle mass
$m$ in the canonical momentum $\vp = m a^2 \dot\vx$ throughout for
brevity.} for canonical coordinate $\vx$.
The temporal and spatial dependences separate at each order: the $i$-th
order growth factor $D_i$ is only a function of scale factor $a$ (or
time $t$), and the $i$-th order displacement field $\vs^{(i)}$ depends
only on $\vq$.
We have also used two types of time derivatives, $\dot\Box \triangleq
\d\Box / \d t$ and $\Box' \triangleq \d\Box / \d \ln a$, related by
$\dot\Box = H \Box'$.

Both the first and second order displacements are potential flows,
\begin{equation}
\vs^{(1)} = - \vnabla_\vq \phi_\vs^{(1)},
\quad
\vs^{(2)} = - \vnabla_\vq \phi_\vs^{(2)},
\end{equation}
with the scalar potentials sourced by
\begin{align}
\nabla_\vq^2 \phi_\vs^{(1)} &= \delta^{(1)}(\vq), \nonumber\\
\nabla_\vq^2 \phi_\vs^{(2)} &= {\textstyle \sum_{i<j}} \bigl(
  \phi_{\vs,ii}^{(1)}\phi_{\vs,jj}^{(1)}
  - \phi_{\vs,ij}^{(1)}\phi_{\vs,ji}^{(1)}
\bigr),
\end{align}
where $\phi_{\vs,ij} \triangleq \p^2\phi_\vs / \p q_i \p q_j$,
and $\delta^{(1)}$ is the linear order of the overdensity field,
$\delta$, related to the density field $\rho$ and mean matter density
$\bar\rho$ by $\delta \triangleq \rho / \bar\rho - 1$.

The linear overdensity $\delta^{(1)}$, which sources the 2LPT particle
initial conditions, is a homogeneous and isotropic Gaussian random field
in the consensus cosmology, with its Fourier transform
$\delta^{(1)}(\vk) = \int\!\d\vq\, \delta^{(1)}(\vq) e^{- i \vk \cdot
\vq}$ characterized by the linear matter power spectrum $\Plin$,
\begin{equation}
\bigl\langle \delta^{(1)}(\vk) \delta^{(1)}(\vk') \bigr\rangle
= (2\pi)^3 \deltaD(\vk + \vk') \Plin(k)
\simeq \deltaK(\vk + \vk') V \Plin(k).
\end{equation}
The angle bracket takes the ensemble average of all possible
realizations.
Homogeneity demands that different wavevectors are uncorrelated, thus
the Dirac delta $\deltaD$ in the first equality.
And with isotropy, $\Plin$ does not depend on the direction of the
wavevector $\vk$, but only on its magnitude, the wavenumber $k
\triangleq |\vk|$.
In a periodic box of volume $V$, where $\vk$ is discrete, $\deltaD$ is
replaced by the Kronecker delta $\deltaK$ in the second equality.
Numerically, we can easily generate a $\delta^{(1)}$ realization by
sampling each Fourier mode independently,
\begin{equation}
\delta^{(1)}(\vk) = \sqrt{V \Plin(k)} \, \omega(\vk),
\end{equation}
with $\omega(\vk)$ being any Hermitian white noise, i.e., Fourier
transform of a real white noise field $\omega(\vq)$.

Cosmological perturbation theory gives the linear power spectrum
as\footnote{Here the time dependence of the linear power, $\Plin(k, a) =
\Plin(k) D_1^2(a)$, has been left to \eqref{lpt}}
\begin{equation}
\frac{k^3}{2\pi^2} \Plin(k)
= \frac{4}{25} \As
  \Bigl( \frac{k}{k_\mathrm{pivot}} \Bigr)^{\ns - 1}
  T^2(k) \frac{c^4 k^4}{\Omegam^2 H_0^4},
\label{Plin}
\end{equation}
where the shape of $\Plin$ is determined by the transfer function $T$,
solution to the linearized Einstein-Boltzmann equations \citep{CAMB,
CLASS}.
$T$ depends on the cosmological parameters,
\begin{equation*}
\vtheta = (
  \As, \ns, \Omegam, \Omegab, h, \cdots),
\end{equation*}
some of which already appear in \eqref{Plin}: $\As$ is the amplitude of
the primordial power spectrum defined at some fixed scale
$k_\mathrm{pivot}$; $\ns$ describes the shape of the primordial power
spectrum; $\Omegam$ is the total matter density parameter; $\Omegab$ is
the baryonic matter density parameter; and $H_0$ is the Hubble constant,
often parameterized by the dimensionless $h$ as $H_0 = 100 h \,
\mathrm{km}/\mathrm{s}/\Mpc$.
Other parameters may enter in extensions of the standard $\Lambda$ cold
dark matter ($\Lambda$CDM) cosmology.

In summary, other than the discretized white noise modes $\vomega$, to
generate initial conditions we need the growth functions $D$ and the
transfer function $T$, both of which depend on the cosmological
parameters $\vtheta$.
We compute $D$ by solving the ordinary differential equations (ODEs)
given in \autoref{app:growth}, and employ the fitting formula for $T$
from \citet{EisensteinHu1998}.
We illustrate these dependencies in the upper left triangle of
\autoref{fig:model}.

At early times and/or lower space/mass resolution, LPT can be accurate
enough to directly compare to the observational data.
However, more expensive integration of the $N$-body dynamics is
necessary in the nonlinear regime.
During LPT and the time integration we can ``observe'' the simulation
predictions by interpolating on the past light cone of a chosen
observer.
These form the upper right square of \autoref{fig:model}.

\vspace{1em}
\subsection{Force Evaluation}
\label{sec:force}

The core of gravitational $N$-body simulation is the gravity solver.
The gravitational potential sourced by matter density fluctuation
satisfies the Poisson equation
\begin{equation}
\nabla^2 \phi(\vx) = \frac32 \frac{\Omegam H_0^2}{a} \delta(\vx),
\end{equation}
where $\nabla^2$ is the Laplacian with respect to $\vx$.
We separate the time dependence by defining $\varphi \triangleq a \phi$,
so $\varphi$ satisfies
\begin{equation}
\nabla^2 \varphi(\vx) = \frac32 \Omegam H_0^2 \, \delta(\vx),
\end{equation}
that only depends on the matter overdensity $\delta$.

While our adjoint method is general, we employ the PM solver in \pmwd\
for efficiency, and leave the implementation of short-range forces to
future development.
With the PM method, we evaluate $\delta(\vx)$ on an auxiliary mesh by
\emph{scattering} particle masses to the nearest grid points.
We use the usual cloud-in-cell (CIC), or trilinear, interpolation
\citep{HockneyEastwood1988}, to compute the fractions of a particle at
$\vx'$ going to a grid point at $\vx$,
\begin{equation}
W(\vx, \vx') = \prod_{i=1}^3
  \max \Bigl( 1 - \frac{|x_i-x'_i|}{l}, 0 \Bigr),
\label{cic}
\end{equation}
where $l$ is the mesh cell size.

The gravitational field, $- \vnabla \varphi$, can then be readily
computed on the mesh with the fast Fourier transform (FFT), as the above
partial differential equation becomes an algebraic one in Fourier space:
\begin{equation}
- k^2 \varphi(\vk) = \frac32 \Omegam H_0^2 \, \delta(\vk).
\end{equation}
In Fourier space, $- \vnabla \varphi(\vx)$ is just $- i \vk
\varphi(\vk)$, each component of which can be transformed back to obtain
the force field.
With 4 (1 forward and 3 inverse) FFTs, we can obtain $- \vnabla
\varphi(\vx)$ from $\delta(\vx)$, with both on the mesh, efficiently.

Finally, we interpolate the particle accelerations by \emph{gathering}
$- \vnabla \varphi$ from the same grid points with the same weights, as
given in \eqref{cic}.

\vspace{1em}
\subsection{Time Integration}
\label{sec:integ}

$N$-body particles move by the following equations of motion
\begin{align}
\dot\vx &= \frac{\vp}{a^2}, \nonumber\\
\dot\vp &= -\vnabla\phi = - \frac{\vnabla\varphi}a.
\label{eom}
\end{align}
We use the \texttt{FastPM} time stepping \citep{FastPM}, designed to
reproduce in the linear regime the linear Lagrangian perturbation
theory, i.e., the 1LPT as the first order in \eqref{lpt}, also known as
the the Zel'dovich approximation (hereafter ZA).
We present a simplified derivation below.

$N$-body simulations integrate \eqref{eom} in discrete steps
(\autoref{fig:loop}), typically with a symplectic integrator that
updates $\vx$ and $\vp$ alternately.
From $t_a$ to $t_b$,
\begin{align}
\vx_b
  &= \vx_a + \int_{t_a}^{t_b} \frac{\vp}{a^2} \,\d t
  \approx \vx_a + \frac{\vp(t_c)}{G_D(t_c)}
    \int_{t_a}^{t_b} \frac{G_D}{a^2} \,\d t
  \triangleq \vx_a + \vp_c D(t_c, t_a, t_b), \nonumber\\
\vp_b
  &= \vp_a - \int_{t_a}^{t_b} \frac{\vnabla\varphi}a \,\d t
  \approx \vp_a - \frac{\vnabla\varphi(t_c)}{G_K(t_c)}
    \int_{t_a}^{t_b} \frac{G_K}a \,\d t
  \triangleq \vp_a - \vnabla\varphi_c K(t_c, t_a, t_b),
\label{eom_integ}
\end{align}
which are named the drift and kick operators, respectively.
In the second equalities of each equation we have introduced two
time-dependent functions $G_D$ and $G_K$.
As approximations, they have been taken out of the integrals together
with $\vp$ and $\vnabla\varphi$ at some intermediate representative time
$t_c$.
We can make the approximation more accurate by choosing $G_D$ to have a
time dependence closer to that of $\vp$.
Likewise for $\vnabla\varphi$ and $G_K$.
However, in most codes $G_D$ and $G_K$ are simply set to 1
\citep{QuinnEtAl1997}, lowering the accuracy when the number of time
steps are limited.

\texttt{FastPM} chooses $G_D$ and $G_K$ according to the ZA growth
history, thereby improving the accuracy on large scales and at early
times.
In ZA, the displacements are proportional to the linear growth factor,
$\vs \propto D_1$, which determines the time dependences of the momenta
and the accelerations by \eqref{eom}.
Therefore, we can set $G_D$ and $G_K$ in \eqref{eom_integ} to
\begin{align}
G_D &:= a^2 \dot D_1, \nonumber\\
G_K &:= a \dot G_D.
\label{G_ZA}
\end{align}
They are functions of $D_1$ and its derivatives, as given by
\eqref{G_eval}.
With these choices, the drift and kick factors, defined in
\eqref{eom_integ}, then become
\begin{align}
D(t_c, t_a, t_b) &=
  \frac{D_1(t_b) - D_1(t_a)}{G_D(t_c)}, \nonumber\\
K(t_c, t_a, t_b) &=
  \frac{G_D(t_b) - G_D(t_a)}{G_K(t_c)}.
\end{align}

While these operators are generally applicable in any symplectic
integrator, we use them in the second order kick-drift-kick leapfrog, or
velocity Verlet, integration scheme \citep{QuinnEtAl1997}.
From $t_{i-1}$ to $t_i$, the particles' state $(\vx, \vp)$ is updated in
the following order:
\begin{alignat}{2}
&K_{i-1}^{i-\half}: &\qquad \vp_{i-\half} &= \vp_{i-1}
  + \va_{i-1} K(t_{i-1}, t_{i-1}, t_{i-\half}), \nonumber\\
&D_{i-1}^i: & \vx_i &= \vx_{i-1}
  + \vp_{i-\half} D(t_{i-\half}, t_{i-1}, t_i), \nonumber\\
&F_i: & \va_i &\triangleq -\vnabla\varphi(\vx_i), \nonumber\\
&K_{i-\half}^i: & \vp_i &= \vp_{i-\half}
  + \va_i K(t_i, t_{i-\half}, t_i).
\label{KDFK}
\end{alignat}
The left column names the operators as shown in \autoref{fig:step}.
The force operator $F$ on the third line computes the accelerations as
described in \autoref{sec:force}.
It caches the results in $\va$, so that they can be used again by the
first $K$ in the next step.
Note that we need to initialize $\va_0$ with $F_0$ before the first time
step.

\vspace{1em}
\subsection{Observation \& Objective}
\label{sec:obsobj}

Because all observables live on our past light cone, we model
observations on the fly by interpolating the $j$-th particle's state
$\hat\vz_j = (\hat\vx_j, \hat\vp_j)$ when they cross the light cone at
$\hat t_j$.
Given $\vz = (\vx, \vp)$ at $t_{i-1}$ and $t_i$, we can parametrize
intermediate particle trajectories with cubic Hermite splines.
Combined with the $\vtheta$-dependent propagation of the light front, we
can solve for the intersections, at which we can record the observed
$\hat\vz$.
The solution can even be analytic if the light propagation is also
approximated cubically.
Note that we only ``observe'' the dark matter phase space here, and
leave more realistic observables to future works, including the forward
modeling of real observational effects.
\autoref{fig:step} illustrates the observation operator $O$ and its
dependencies on the previous and the current time step.

We can compare the simulated observables to the observational data,
either at level of the fields or the summary statistics, by some
objective function in case of optimization, or by a posterior
probability for Bayesian inference.
We refer to both cases by objective and denote it by $\cJ$ throughout.
Note that in general it can also depend on $\vtheta$ and $\vomega$, in
the form of regularization or prior probability.

Formally, we can combine the observation and objective operators as
\begin{equation}
\cJ(\hat\vz, \vtheta, \vomega)
  = \cJ \bigl(
    \hat\vz(\vz_0, \cdots, \vz_n, \vtheta), \vtheta, \vomega
  \bigr),
\label{O}
\end{equation}
as illustrated in the lower right triangle in \autoref{fig:model}.
Note this form also captures the conventional simulation snapshots at
the end of a time step or those interpolated between two consecutive
steps, so we model all these cases as observations in \pmwd.

\vspace{1em}
\section{Backward Differentiation -- the Adjoint Method}
\label{sec:bwd}

We first introduce the adjoint method for generic time-dependent ODEs,
and derive the adjoint equations following a pedagogical tutorial by
\citet{Bradley2019}.
We then adopt the discretize-then-optimize approach, and derive the
discrete adjoint equations, that is more suitable for the $N$-body
symplectic time integration.
Finally we apply them to derive the adjoint equations and the gradients
for cosmological simulations described in \autoref{sec:fwd}, and couple
them with the reverse time integration to reduce the space complexity.

\vspace{1em}
\subsection{Variational (Tangent) and Adjoint Equations}

Consider a vector state $\vz(t)$ subject to the following ODEs and
initial conditions
\begin{equation}
\dot\vz = \vf(\vz, t, \vtheta),
\qquad
\vz(t_0) = \vz_0(\vtheta),
\label{ode}
\end{equation}
for $t \in [t_0, t_1]$.
Here the initial conditions can depend on the parameters $\vtheta$.

A perturbation in the initial conditions propagates forward in time.
For $\vz(t, t_0, \vz_0)$, the Jacobian of state variables describing
this,
\begin{equation}
\vDelta = \frac{\p\vz}{\p\vz_0},
\end{equation}
evolves from identity $\vDelta_0 = \vI$ by
\begin{equation}
\dot\vDelta = \frac{\p\vf}{\p\vz} \cdot \vDelta,
\label{variational}
\end{equation}
following from \eqref{ode}.
\eqref{variational} is known as the \emph{variational or tangent
equation}.

The backward version of $\vDelta$,
\begin{equation}
\vLambda = \frac{\p\vz_1}{\p\vz},
\end{equation}
evolves backward in time from identity $\vLambda_1 = \vI$ by
\begin{equation}
\dot\vLambda = - \vLambda \cdot \frac{\p\vf}{\p\vz}.
\label{adjoint}
\end{equation}
\eqref{adjoint} is called the \emph{adjoint equation}, for the
right-hand side is $(\p\vf / \p\vz)^\intercal \vLambda$.
It can be derived from the time-invariance of $\vLambda \cdot \vDelta$:
\begin{equation}
\vLambda \cdot \dot\vDelta + \dot\vLambda \cdot \vDelta
= \frac\d{\d t} (\vLambda \cdot \vDelta)
= \frac\d{\d t} \Bigl( \frac{\p\vz_1}{\p\vz}
  \cdot \frac{\p\vz}{\p\vz_0} \Bigr)
= \frac\d{\d t} \frac{\p\vz_1}{\p\vz_0}
= \vzero.
\end{equation}

Alternatively, the adjoint equation can be derived from the variational
equation, using the facts that $\d \bm{M}^{-1} / \d t = - \bm{M}^{-1} \,
\dot{\bm{M}} \, \bm{M}^{-1}$ for any invertible matrix $\bm{M}$, and
$(\p\vz / \p\tilde\vz)^{-1} = \p\tilde\vz / \p\vz$.
Like \eqref{variational} and \eqref{adjoint}
\begin{equation}
\frac\d{\d t} \frac{\p\vz}{\p\tilde\vz}
= \frac{\p\vf}{\p\vz} \cdot \frac{\p\vz}{\p\tilde\vz},
\qquad
\frac\d{\d t} \frac{\p\tilde\vz}{\p\vz}
= - \frac{\p\tilde\vz}{\p\vz} \cdot \frac{\p\vf}{\p\vz}.
\end{equation}
As we can see next, the adjoint equation takes a similar form when one
optimizes an objective function of the state.

\vspace{1em}
\subsection{Objective on the Final State}

In the simplest case, the objective function depends only on the final
state, e.g., the last snapshot of a simulation, and possibly the
parameters too in the form of regularization or prior information, i.e.,
$\cJ(\vz_1, \vtheta)$.
To optimize the objective under the constraint given by the ODEs, we can
introduce a time-dependent function $\vlambda(t)$ as the Lagrange
multiplier:
\begin{equation}
\cL = \cJ(\vz_1, \vtheta)
- \int_{t_0}^{t_1} \vlambda(t)
  \cdot \bigl[ \dot\vz - \vf(\vz, t, \vtheta) \bigr] \, \d t.
\end{equation}
Note the minus sign we have introduced in front of $\vlambda$ for later
convenience.

Its total derivative with respect to $\vtheta$ is
\begin{equation}
\frac{\d\cL}{\d\vtheta}
= \frac{\p\cJ}{\p\vtheta}
+ \frac{\p\cJ}{\p\vz_1} \cdot \frac{\p\vz_1}{\p\vtheta}
- \int_{t_0}^{t_1} \vlambda
  \cdot \Bigl( \frac{\p\dot\vz}{\p\vtheta}
    - \frac{\p\vf}{\p\vz} \cdot \frac{\p\vz}{\p\vtheta}
    - \frac{\p\vf}{\p\vtheta} \Bigr) \, \d t.
\end{equation}
Integrating the first term of the integrand by parts:
\begin{equation}
\int_{t_0}^{t_1} \vlambda \cdot \frac{\p\dot\vz}{\p\vtheta} \, \d t
= \int_{t_0}^{t_1} \vlambda \cdot \frac{\p^2\vz}{\p\vtheta\p t} \, \d t
= \Bigl[ \vlambda \cdot \frac{\p\vz}{\p\vtheta} \Bigr]_{t_0}^{t_1}
- \int_{t_0}^{t_1} \dot \vlambda \cdot \frac{\p\vz}{\p\vtheta} \, \d t,
\end{equation}
and plugging it back:
\begin{equation}
\frac{\d\cL}{\d\vtheta}
= \frac{\p\cJ}{\p\vtheta}
+ \Bigl( \frac{\p\cJ}{\p\vz_1} - \vlambda_1 \Bigr)
  \cdot \frac{\p\vz_1}{\p\vtheta}
+ \vlambda_0 \cdot \frac{\p\vz_0}{\p\vtheta}
+ \int_{t_0}^{t_1} \Bigl[
  \Bigl( \vlambda \cdot \frac{\p\vf}{\p\vz} + \dot\vlambda \Bigr)
  \cdot \frac{\p\vz}{\p\vtheta}
  + \vlambda \cdot \frac{\p\vf}{\p\vtheta} \Bigr] \, \d t.
\end{equation}

Now we are free to choose
\begin{equation}
\dot\vlambda = - \vlambda \cdot \frac{\p\vf}{\p\vz},
\qquad
\vlambda_1 = \frac{\p\cJ}{\p\vz_1},
\label{adjoint1}
\end{equation}
that allow us to avoid all $\p\vz/\p\vtheta$ terms in the final
objective gradient:
\begin{equation}
\frac{\d\cJ}{\d\vtheta}
= \frac{\d\cL}{\d\vtheta}
= \frac{\p\cJ}{\p\vtheta} + \vlambda_0 \cdot \frac{\p\vz_0}{\p\vtheta}
+ \int_{t_0}^{t_1} \vlambda \cdot \frac{\p\vf}{\p\vtheta} \, \d t,
\end{equation}
in which the first two terms come from the regularization and initial
conditions, respectively.
\eqref{adjoint1} is the adjoint equation for the objective $\cJ(\vz_1,
\vtheta)$.
With the initial conditions set at the final time, we can integrate it
backward in time to obtain $\vlambda(t)$, which enters the above
equation and yields $\d\cJ / \d\vtheta$.

Note that \eqref{adjoint1} has the same form as \eqref{adjoint}, and
their solutions are related by
\begin{equation}
\vlambda = \vlambda_1 \cdot \vLambda
= \frac{\p\cJ}{\p\vz_1} \cdot \frac{\p\vz_1}{\p\vz}
= \frac{\p\cJ}{\p\vz}.
\end{equation}
And like $\vLambda \cdot \vDelta$, $\vlambda \cdot \vDelta$ is
time-invariant:
\begin{equation}
\vlambda_0 \cdot \vDelta_0
= \vlambda_1 \cdot \vDelta_1
= \vlambda \cdot \vDelta
= \frac{\p\cJ}{\p\vz} \cdot \frac{\p\vz}{\p\vz_0}
= \frac{\p\cJ}{\p\vz_0}
= \vlambda_0.
\end{equation}
Computing $\vlambda_1 \cdot \vDelta_1$ directly is expensive, but
solving \eqref{adjoint1} backward for $\vlambda_0$ is cheap.
This is related to the fact that the reverse-mode AD or backpropagation
is cheaper than the forward mode for optimization.

\vspace{1em}
\subsection{Objective on the State History}

The adjoint method applies to more complex cases too.
Let's consider an objective as a functional of the evolution history
with some regularization $\cR$ on $\vtheta$
\begin{equation}
\cJ = \cR(\vtheta) + \int_{t_0}^{t_1} g(\vz, t, \vtheta) \, \d t.
\end{equation}
The derivation is similar:
\begin{equation}
\cL = \cR(\vtheta)
+ \int_{t_0}^{t_1} \bigl\{ g(\vz, t, \vtheta) -
  \vlambda(t) \cdot \bigl[ \dot\vz - \vf(\vz, t, \vtheta) \bigr]
  \bigr\} \, \d t
\end{equation}
has gradient
\begin{align}
\frac{\d\cL}{\d\vtheta}
&= \frac{\d\cR}{\d\vtheta}
+ \int_{t_0}^{t_1} \Bigl[
    \frac{\p g}{\p\vz} \cdot \frac{\p\vz}{\p\vtheta}
    + \frac{\p g}{\p\vtheta}
    - \vlambda \cdot \Bigl( \frac{\p\dot\vz}{\p\vtheta}
    - \frac{\p\vf}{\p\vz} \cdot \frac{\p\vz}{\p\vtheta}
    - \frac{\p\vf}{\p\vtheta} \Bigr) \Bigr] \, \d t \\
&= \frac{\d\cR}{\d\vtheta}
- \vlambda_1 \cdot \frac{\p\vz_1}{\p\vtheta}
+ \vlambda_0 \cdot \frac{\p\vz_0}{\p\vtheta}
+ \int_{t_0}^{t_1} \Bigl[
  \Bigl( \frac{\p g}{\p\vz} + \vlambda \cdot \frac{\p\vf}{\p\vz}
    + \dot\vlambda \Bigr) \cdot \frac{\p\vz}{\p\vtheta}
  + \frac{\p g}{\p\vtheta} + \vlambda \cdot \frac{\p\vf}{\p\vtheta}
\Bigr] \, \d t. \nonumber
\end{align}

The adjoint equation becomes
\begin{equation}
\dot\vlambda = - \frac{\p g}{\p\vz}
  - \vlambda \cdot \frac{\p\vf}{\p\vz},
\qquad
\vlambda_1 = \vzero,
\end{equation}
with the objective gradient
\begin{equation}
\frac{\d\cJ}{\d\vtheta}
= \frac{\d\cL}{\d\vtheta}
= \frac{\d\cR}{\d\vtheta} + \vlambda_0 \cdot \frac{\p\vz_0}{\p\vtheta}
+ \int_{t_0}^{t_1} \Bigl( \frac{\p g}{\p\vtheta}
  + \vlambda \cdot \frac{\p\vf}{\p\vtheta} \Bigr) \, \d t.
\end{equation}

The time-invariant is now
\begin{equation}
\vlambda_1 \cdot \vDelta_1
= \vlambda_0 \cdot \vDelta_0
  + \int_{t_0}^{t_1} \frac{\p g}{\p\vz} \cdot \vDelta \, \d t
= \vlambda \cdot \vDelta
  + \int_t^{t_1} \frac{\p g}{\p\vz} \cdot \vDelta \, \d t'
= \vzero,
\end{equation}
so
\begin{equation}
\vlambda \cdot \vDelta
= - \int_t^{t_1} \frac{\p g}{\p\vz} \cdot \vDelta \, \d t'
= - \int_t^{t_1} \frac{\p g}{\p\vz_0} \, \d t'.
\end{equation}

\vspace{1em}
\subsection{Objective on the Observables}
\label{sec:lightcone}

Now let's consider an objective that depends on different state
components at different times, e.g., in the Universe where further
objects intersected our past light cone earlier.
It falls between the previous two scenarios, and we can derive its
adjoint equation similarly.

We denote the observables by $\hat\vz$, with different components $\hat
z_j$ affecting the objective $\cJ(\hat\vz, \vtheta)$ at different times
$\hat t_j$, i.e., $\hat z_j \triangleq z_j(\hat t_j)$.
The Lagrangian becomes
\begin{equation}
\cL = \cJ(\hat\vz, \vtheta)
- \sum_j \int_{t_0}^{\hat t_j} \lambda_j(t)
  \bigl[ \dot z_j - f_j(\vz, t, \vtheta) \bigr] \, \d t,
\end{equation}
constraining only the parts of trajectories inside the light cone.
Its gradient is
\begin{align}
\frac{\d\cL}{\d\vtheta}
&= \frac{\p\cJ}{\p\vtheta}
+ \frac{\p\cJ}{\p\hat\vz} \cdot \frac{\p\hat\vz}{\p\vtheta}
- \sum_j \int_{t_0}^{\hat t_j}
  \Bigl( \lambda_j \frac{\p\dot z_j}{\p\vtheta}
    - \vlambda \cdot \frac{\p\vf}{\p z_j} \frac{\p z_j}{\p\vtheta}
    - \lambda_j \frac{\p f_j}{\p\vtheta} \Bigr) \, \d t \\
&= \frac{\p\cJ}{\p\vtheta}
+ \Bigl( \frac{\p\cJ}{\p\hat\vz} - \hat\vlambda \Bigr)
  \cdot \frac{\p\hat\vz}{\p\vtheta}
+ \vlambda_0 \cdot \frac{\p\vz_0}{\p\vtheta}
+ \sum_j \int_{t_0}^{\hat t_j} \Bigl[
  \Bigl( \vlambda \cdot \frac{\p\vf}{\p z_j} + \dot\lambda_j \Bigr)
  \frac{\p z_j}{\p\vtheta}
  + \lambda_j \frac{\p f_j}{\p\vtheta} \Bigr] \, \d t, \nonumber
\end{align}
where we have defined $\hat\vlambda$ similarly with components
$\hat\lambda_j \triangleq \lambda_j(\hat t_j)$.
In the first equality, we have also dropped a vanishing term $\sum_j
\hat\lambda_j \bigl[ \dot z_j - f_j(\hat t_j) \bigr] \p\hat t_j /
\p\vtheta$, i.e., $\p\hat t_j / \p\vtheta$ does not directly enter the
gradient.

Now we find the adjoint equation
\begin{equation}
\dot\vlambda = - \vlambda \cdot \frac{\p\vf}{\p\vz},
\qquad
\hat\vlambda = \frac{\p\cJ}{\p\hat\vz},
\end{equation}
that has the same form as \eqref{adjoint1}, with a slightly different
initial condition given at the respective observation time of each
component.
The objective gradient is also similar to the previous cases:
\begin{equation}
\frac{\d\cJ}{\d\vtheta}
= \frac{\d\cL}{\d\vtheta}
= \frac{\p\cJ}{\p\vtheta} + \vlambda_0 \cdot \frac{\p\vz_0}{\p\vtheta}
+ \sum_j \int_{t_0}^{\hat t_j} \lambda_j \frac{\p f_j}{\p\vtheta} \, \d t.
\end{equation}

Note that even though $\lambda_j(t)$ for $t > \hat t_j$ does not affect
the final gradient, they can enter the right-hand side of the adjoint
equation, and affect those $\lambda_k$ with $t < \hat t_k$, i.e., inside
the light cone.
Physically however, $\p f_j / \p z_k$ should vanish for spacelike
separated pairs of $z_j$ and $z_k$, even though the Newtonian
approximation we adopt introduces some small deviation.
Therefore, we can set $\lambda_j(t)$ to 0 for $t > \hat t_j$, and bump
it to $\p\cJ / \p\hat z_j$ at $\hat t_j$.

\vspace{1em}
\subsection{Discretize Then Optimize}

In practice, the time integration of \eqref{ode} is discrete.
Consider the explicit methods,
\begin{equation}
\vz_{i+1} = \vz_i + \vF_i(\vz_i, \vtheta), \qquad i = 0, \cdots, n-1,
\label{explicit}
\end{equation}
which include the leapfrog integrator commonly used for Hamiltonian
dynamics.
We want to propagate the gradients backward along the \emph{same}
discrete trajectory as taken by the forward integration.
Therefore, instead of the continuous adjoint equations derived above, we
need the adjoint method for the discrete integrator.

Without loss of generality, we derive the adjoint equation for an
objective depending on the state at all time steps, which can be easily
specialized to the 3 cases discussed above with only slight
modifications.
The discretized Lagrangian is now
\begin{equation}
\cL = \cJ(\vz_0, \cdots, \vz_n, \vtheta)
- \sum_{i=0}^{n-1} \vlambda_{i+1}
  \cdot \bigl[ \vz_{i+1} - \vz_i - \vF_i(\vz_i, \vtheta) \bigr],
\end{equation}
whose gradient is
\begin{align}
\frac{\d\cL}{\d\vtheta}
&= \frac{\p\cJ}{\p\vtheta}
+ \sum_{i=0}^n \frac{\p\cJ}{\p\vz_i} \cdot \frac{\p\vz_i}{\p\vtheta}
- \sum_{i=0}^{n-1} \vlambda_{i+1}
  \cdot \Bigl( \frac{\p\vz_{i+1}}{\p\vtheta} - \frac{\p\vz_i}{\p\vtheta}
    - \frac{\p\vF_i}{\p\vz_i} \cdot \frac{\p\vz_i}{\p\vtheta}
    - \frac{\p\vF_i}{\p\vtheta} \Bigr) \nonumber\\
&= \frac{\p\cJ}{\p\vtheta}
+ \sum_{i=0}^n \frac{\p\cJ}{\p\vz_i} \cdot \frac{\p\vz_i}{\p\vtheta}
- \sum_{i=1}^n \vlambda_i \cdot \frac{\p\vz_i}{\p\vtheta}
+ \sum_{i=0}^{n-1} \vlambda_{i+1}
  \cdot \Bigl( \frac{\p\vz_i}{\p\vtheta}
    + \frac{\p\vF_i}{\p\vz_i} \cdot \frac{\p\vz_i}{\p\vtheta}
    + \frac{\p\vF_i}{\p\vtheta} \Bigr) \nonumber\\
&= \frac{\p\cJ}{\p\vtheta}
+ \Bigl( \frac{\p\cJ}{\p\vz_n} - \vlambda_n \Bigr)
  \cdot \frac{\p\vz_n}{\p\vtheta}
+ \vlambda_0 \cdot \frac{\p\vz_0}{\p\vtheta} \nonumber\\
&\hspace{4em} + \sum_{i=0}^{n-1} \Bigl[
  \Bigl( \frac{\p\cJ}{\p\vz_i} - \vlambda_i + \vlambda_{i+1}
    + \vlambda_{i+1} \cdot \frac{\p\vF_i}{\p\vz_i} \Bigr)
  \cdot \frac{\p\vz_i}{\p\vtheta}
  + \vlambda_{i+1} \cdot \frac{\p\vF_i}{\p\vtheta} \Bigr].
\end{align}

So the discrete adjoint equation is
\begin{equation}
\vlambda_{i-1} = \vlambda_i
  + \vlambda_i \cdot \frac{\p\vF_{i-1}}{\p\vz_{i-1}}
  + \frac{\p\cJ}{\p\vz_{i-1}},
\quad
i = n, \cdots, 1;
\qquad
\vlambda_n = \frac{\p\cJ}{\p\vz_n}.
\label{explicit_adj}
\end{equation}
We can iterate it backward in time to compute the final objective
gradient:
\begin{equation}
\frac{\d\cJ}{\d\vtheta}
= \frac{\d\cL}{\d\vtheta}
= \frac{\p\cJ}{\p\vtheta} + \vlambda_0 \cdot \frac{\p\vz_0}{\p\vtheta}
+ \sum_{i=1}^n \vlambda_i \cdot \frac{\p\vF_{i-1}}{\p\vtheta}.
\label{explicit_grad}
\end{equation}
These equations are readily adaptable to simulated observables.
For snapshots at $t_n$ or interpolated between $t_{n-1}$ and $t_n$,
all $\p\cJ / \p\vz_i$ vanish except for the last one or two,
respectively.
For light cones, as discussed in \autoref{sec:lightcone}, each component
of $\hat\vz$ is interpolated at different times, thus all $\p\cJ /
\p\vz_i$ vanish except for those times relevant for its interpolation,
and the corresponding $\lambda_i$ can be set to zero for $i$ greater
than the intersection time.

At the $i$-th iteration the adjoint variable requires the
vector-Jacobian product (VJP) $\vlambda_i \cdot \p\vF_{i-1} /
\p\vz_{i-1}$ and the partial objective derivative $\p\cJ / \p\vz_{i-1}$
at the next time step, which can be easily computed by AD if the whole
forward history of \eqref{explicit} has been saved.
However, this can be extremely costly in memory, which can be alleviated
by checkpointing algorithms such as Revolve and its successors
\citep{Revolve}.
Alternatively, if a solution to \eqref{ode} is unique, we can integrate
it backward and recover the history, which is easy for reversible
Hamiltonian dynamics and with reversible integrators such as leapfrog.
When the $N$-body dynamics become too chaotic, one can use more precise
floating-point numbers and/or save multiple checkpoints\footnote{This is
different from the checkpointing in the Revolve algorithm, which needs
to rerun the forward iterations.} during the forward evolution, from
which the backward evolution can be resumed piecewise.

\vspace{1em}
\subsection{Application to Simulation}
\label{sec:adj}

The adjoint method provides systematic ways of deriving the objective
gradient under constraints \citep{Pontryagin1962}, here imposed by the
$N$-body equations of motion.
We have introduced above the adjoint method for generic time-dependent
problems in both continuous and discrete cases.
The continuous case is easier to understand and has pedagogical values,
while the discrete case is the useful one in our application, for we
want to propagate numerically the gradients backward along the
\emph{same} path as that of the forward time integration.

For the $N$-body particles, the state variable\footnote{State and
adjoint vectors in the adjoint equations include enumeration of
particles, e.g., $\vnabla\vvarphi$ includes the $\vnabla\varphi$ of each
particle.} is $\vz = (\vx, \vp)^\intercal$.
Their adjoint variables help to accumulate the objective gradient while
evolving backward in time by the adjoint equation.
Let's denote them by $\vlambda = (\vxi, \vpi)$.
We can compare each step of \eqref{KDFK} and \eqref{O} to
\eqref{explicit}, and write down its adjoint equation following
\eqref{explicit_adj}.
Taking $D_{i-1}^i$ for example, we can write it explicitly as
\begin{equation*}
\begin{pmatrix} \vx_i \\ \vp_{i-\half} \end{pmatrix}
= \begin{pmatrix} \vx_{i-1} \\ \vp_{i-\half} \end{pmatrix}
+ \begin{pmatrix}
    \vp_{i-\half} D(t_{i-\half}, t_{i-1}, t_i) \\ \vzero
  \end{pmatrix},
\end{equation*}
in the form of \eqref{explicit}.
By \eqref{explicit_adj}, its adjoint equation is
\begin{align*}
\begin{pmatrix} \vxi_{i-\half} & \vpi_{i-1} \end{pmatrix}
&= \begin{pmatrix} \vxi_{i-\half} & \vpi_i \end{pmatrix}
+ \begin{pmatrix} \vxi_{i-\half} & \vpi_i \end{pmatrix}
  \begin{pmatrix}
    \vzero & \vI \, D(t_{i-\half}, t_{i-1}, t_i) \\
    \vzero & \vzero \\
  \end{pmatrix} \\
&= \begin{pmatrix} \vxi_{i-\half} & \vpi_i \end{pmatrix}
- \begin{pmatrix}
    \vzero & \vxi_{i-\half} D(t_{i-\half}, t_i, t_{i-1})
  \end{pmatrix},
\end{align*}
where we have used the fact that $D(t_c, t_a, t_b) = - D(t_c, t_b,
t_a)$, and left the $\p\cJ / \p\vz_i$ term, the explicit dependence of
the objective on the intermediate states (from, e.g., observables on the
light cone), to the observation operator $O$ below.
This also naturally determines the subscripts of $\xi$ and $\pi$.

Repeating the derivation for $K$ and $O$, and flipping the arrow of
time, we present the adjoint equation time stepping for \eqref{KDFK}
from $t_i$ to $t_{i-1}$:
\begin{alignat}{2}
&K_i^{i-\half}: &\qquad \vp_{i-\half} &= \vp_i
  + \va_i K(t_i, t_i, t_{i-\half}), \nonumber\\
&& \vxi_{i-\half} &= \vxi_i
  - \valpha_i K(t_i, t_i, t_{i-\half}), \nonumber\\
&D_i^{i-1}: & \vx_{i-1} &= \vx_i
  + \vp_{i-\half} D(t_{i-\half}, t_i, t_{i-1}), \nonumber\\
&& \vpi_{i-1} &= \vpi_i
  - \vxi_{i-\half} D(t_{i-\half}, t_i, t_{i-1}), \nonumber\\
&F_{i-1}: & \va_{i-1} &= -\vnabla\vvarphi(\vx_{i-1}), \nonumber\\
&& \valpha_{i-1} &\triangleq - \vpi_{i-1}
  \cdot \frac{\p\vnabla\vvarphi_{i-1}}{\p\vx_{i-1}}, \nonumber\\
&& \vzeta_{i-1} &\triangleq - \vpi_{i-1}
  \cdot \frac{\p\vnabla\vvarphi_{i-1}}{\p\vtheta}, \nonumber\\
&K_{i-\half}^{i-1}: & \vp_{i-1} &= \vp_{i-\half}
  + \va_{i-1} K(t_{i-1}, t_{i-\half}, t_{i-1}), \nonumber\\
&& \vxi_{i-1} &= \vxi_{i-\half}
  - \valpha_{i-1} K(t_{i-1}, t_{i-\half}, t_{i-1}), \nonumber\\
&O_i^{i-1}: &\qquad \vxi_{i-1} :\:\!\!\!&= \vxi_{i-1}
  + \frac{\p\cJ}{\p\vx_{i-1}}, \nonumber\\
&& \vpi_{i-1} :\:\!\!\!&= \vpi_{i-1} + \frac{\p\cJ}{\p\vp_{i-1}}.
\label{KDFKO_adj}
\end{alignat}
Like $\va$, we have introduced $\valpha$ and $\vzeta$ to cache the
vector-Jacobian products on their right-hand sides, for the next time
step in the kick operator and the objective gradient (see below),
respectively.
Note that in the reverse order, the $F$ operator is at $t_{i-1}$ instead
of $t_i$ as in \eqref{KDFK}, and we need to initialize $\va_n$,
$\valpha_n$, and $\vzeta_n$ with $F_n$ before stepping from $t_n$ to
$t_{n-1}$.
Likewise, the gradient of $O_n^{n-1}$ at $t_n$ is absent in
\eqref{KDFKO_adj} but enters via the initial conditions following
\eqref{explicit_adj}.
Explicitly, the initial conditions of \eqref{KDFKO_adj} are
\begin{equation}
\vxi_n = \frac{\p\cJ}{\p\vx_n},
\vpi_n = \frac{\p\cJ}{\p\vp_n},
\va_n = -\vnabla\vvarphi(\vx_n),
\valpha_n = - \vpi_n \cdot \frac{\p\vnabla\vvarphi_n}{\p\vx_n},
\vzeta_n = - \vpi_n \cdot \frac{\p\vnabla\vvarphi_n}{\p\vtheta}.
\end{equation}

The VJPs in $F$ and the $\p\cJ / \p\vz$'s in $O$ can be computed by AD
if the whole forward integration and observation history of \eqref{KDFK}
and \eqref{O} has been saved.
However, this can be too costly spatially for GPUs, whose memories are
much smaller than those of CPUs.
Alternatively, we take advantage of the reversibility of the $N$-body
dynamics and the leapfrog integrator, and recover the history by reverse
time integration, which we have already included on the first lines of
the $K$ and $D$ operators in \eqref{KDFKO_adj}.
We can integrate the leapfrog and the adjoint equations jointly backward
in time, and still benefit from the convenience of AD in computing VJPs
and $\p\cJ / \p\vz$'s.
In practice, the numerical reversibility suffers from the finite
precision and the chaotic $N$-body dynamics, which we find is generally
not a concern for our applications in the result section.

Finally, during the reverse time integration, we can accumulate the
objective gradient following \eqref{explicit_grad}:
\begin{align}
\frac{\d\cJ}{\d\vtheta} =& \frac{\p\cJ}{\p\vtheta}
  + \frac{\p\cJ}{\p\hat\vx} \cdot \frac{\p\hat\vx}{\p\vtheta}
  + \frac{\p\cJ}{\p\hat\vp} \cdot \frac{\p\hat\vp}{\p\vtheta}
  + \vxi_0 \cdot \frac{\p\vx_0}{\p\vtheta}
  + \vpi_0 \cdot \frac{\p\vp_0}{\p\vtheta} \nonumber\\
  &- \sum_{i=1}^n \Bigl[
    \bigl( \vpi_i \cdot\va_i \frac\p{\p\vtheta} + \vzeta_i
      \bigr) K(t_i, t_i, t_{i-\half})
    + \vxi_{i-\half} \cdot \vp_{i-\half}
      \frac{\p D(t_{i-\half}, t_i, t_{i-1})}{\p\vtheta} \nonumber\\
    &\qquad + \bigl( \vpi_{i-1} \cdot \va_{i-1} \frac\p{\p\vtheta}
      + \vzeta_{i-1} \bigr) K(t_{i-1}, t_{i-\half}, t_{i-1})
    \Bigr], \nonumber\\
\frac{\d\cJ}{\d\vomega} =& \frac{\p\cJ}{\p\vomega}
  + \vxi_0 \cdot \frac{\p\vx_0}{\p\vomega}
  + \vpi_0 \cdot \frac{\p\vp_0}{\p\vomega},
\label{KDFKO_grad}
\end{align}
where the latter backpropagates from fewer sources than the former as
shown in \autoref{fig:model}.
To implement \eqref{KDFKO_adj}-\eqref{KDFKO_grad} in \pmwd\ with
\texttt{JAX}, we only need to write custom VJP rules for the high-level
$N$-body integration-observation loop, while the derivatives and VJPs of
the remaining parts including regularization/prior, the observation, the
initial conditions, the kick and drift factors, the growth and transfer
functions, etc., can all be conveniently computed by AD.

Other than that, we also implement custom VJPs for the scatter and
gather operations in \autoref{sec:force} following \citet{Feng2017}, and
these further save memory in gradient computations of those nonlinear
functions.

\begin{figure*}[t]
\centering
\includegraphics[width=0.85\linewidth]{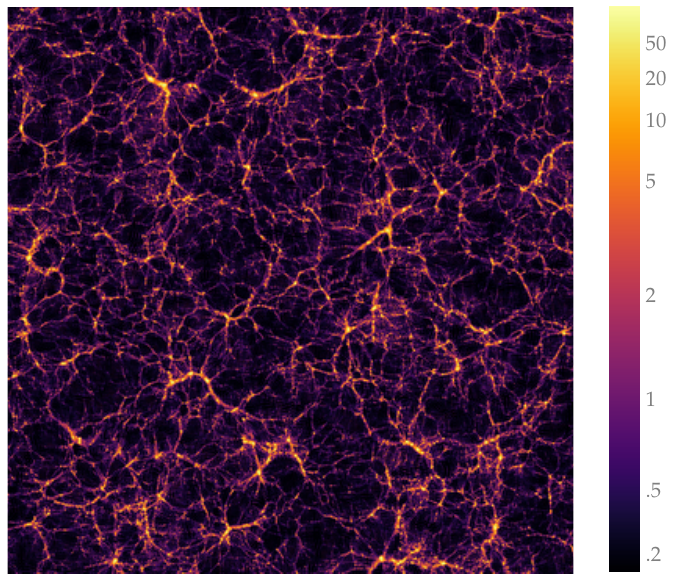}
\caption{Relative matter density field, $1+\delta$, at $a=1$, projected
from an $8\,\Mpc/h$ thick slab in a \pmwd\ simulation, that has evolved
$512^3$ particles with single precision and a $1024^3$ mesh in a
$(512\,\Mpc/h)^3$ box for 63 time steps.
The simulation takes only 13 seconds to finish on an \GPU\ GPU.
}
\label{fig:slab}
\end{figure*}

\vspace{1em}
\section{Results}

We implement the forward simulation and backward adjoint method in the
PM code \pmwd.
All results assume a simple $\Lambda$CDM cosmology: $\vtheta = (\As =
2\times10^{-9}, \ns = 0.96, \Omegam = 0.3, \Omegab = 0.05, h = 0.7)$,
and use an \GPU\ GPU with 80GB memory.

As in \texttt{FastPM}, the choice of time steps is flexible, and in fact
with \pmwd\ it can even be optimized in non-parametric ways to improve
simulation accuracy at fixed computation cost.
Here we use time steps linearly spaced in scale factor $a$, and leave
such optimization in a follow-up work.

\vspace{1em}
\subsection{Simulation}

We first test the forward simulations of \pmwd.
\autoref{fig:slab} shows the cosmic web in the final snapshot of a
fairly large simulation for the size of GPU memories.

Because GPUs are inherently parallel devices, they can output different
results for identical inputs.
To test the reproducibility, in \autoref{tab:rep} we compare the
root-mean-square deviations (RMSDs) of particle displacements and
velocities between two runs, relative to their respective standard
deviations, with different floating-point precisions, mesh sizes,
particle masses, and numbers of time steps.
Other than the precision, mesh size is the most important factor,
because a finer mesh can better resolve the most nonlinear and dense
structures, which can affect reproducibility as the order of many
operations can change easily.
The particle mass plays a similar role and less massive particles
generally take part in nonlinear motions at earlier times.
The number of time steps has very small impact except in the most
nonlinear cases.
And interestingly, more time steps improves the reproducibility most of
the time.

\begin{table*}[t]
\centering
\caption{\pmwd\ reproducibility on GPU.
GPUs can output different results for identical inputs.
We simulate $384^3$ particles from $a=1/64$ to $a=1$, with two
floating-point precisions, two mesh sizes, two particle masses (with box
sizes of $192 \, \Mpc / h$ and $384 \, \Mpc / h$), and two time step
sizes.
We take the root-mean-square deviations (RMSDs) of particle
displacements and velocities between two runs at $a=1$, and quote their
ratios to the respective standard deviations, about $6 \, \Mpc / h$ and
$3\times10^2 \, \mathrm{km} / \mathrm{s}$.
In general, the factors on the left of the left four columns affect the
reproducibility more than the those on the right, and lower rows are
more reproducible than the upper ones.
}
\label{tab:rep}
\begin{tabular}{cccccc}
  \toprule
  precision & cell/ptcl & ptcl mass $[10^{10} M_\odot]$ & time steps
    & disp rel diff & vel rel diff \\
  \midrule
  single & 8 & 1 &  63 & $1.6\times10^{-6}$  & $4.8\times10^{-5}$  \\
  single & 8 & 1 & 126 & $5.1\times10^{-7}$  & $1.3\times10^{-5}$  \\
  single & 8 & 8 &  63 & $3.1\times10^{-7}$  & $2.4\times10^{-6}$  \\
  single & 8 & 8 & 126 & $3.1\times10^{-7}$  & $2.3\times10^{-6}$  \\
  single & 1 & 1 &  63 & $1.4\times10^{-7}$  & $1.5\times10^{-6}$  \\
  single & 1 & 1 & 126 & $1.3\times10^{-7}$  & $1.4\times10^{-6}$  \\
  single & 1 & 8 &  63 & $1.1\times10^{-7}$  & $4.5\times10^{-7}$  \\
  single & 1 & 8 & 126 & $9.9\times10^{-8}$  & $4.4\times10^{-7}$  \\
  double & 8 & 1 &  63 & $1.4\times10^{-15}$ & $3.6\times10^{-14}$ \\
  double & 8 & 1 & 126 & $8.3\times10^{-16}$ & $1.7\times10^{-14}$ \\
  double & 8 & 8 &  63 & $5.4\times10^{-16}$ & $4.1\times10^{-15}$ \\
  double & 8 & 8 & 126 & $5.9\times10^{-16}$ & $4.3\times10^{-15}$ \\
  double & 1 & 1 &  63 & $2.4\times10^{-16}$ & $2.3\times10^{-15}$ \\
  double & 1 & 1 & 126 & $2.3\times10^{-16}$ & $2.3\times10^{-15}$ \\
  double & 1 & 8 &  63 & $1.9\times10^{-16}$ & $7.9\times10^{-16}$ \\
  double & 1 & 8 & 126 & $1.8\times10^{-16}$ & $7.8\times10^{-16}$ \\
  \bottomrule
\end{tabular}
\end{table*}

\vspace{1em}
\subsection{Differentiation}

Model differentiation evolves the adjoint equations backward in time.
To save memory, the trajectory of the model state in the forward run is
not saved, but re-simulated together with the adjoint equations.
Even though in principle the $N$-body systems are reversible, in
practice the reconstructed trajectory can differ from the forward one
due to the finite numerical precision and exacerbated by the chaotic
dynamics.
Better reversibility means the gradients propagate backward along a
trajectory closer to the forward path, and thus would be more accurate.
To test this, in \autoref{tab:rev} we compare the RMSDs of the
forward-then-reverse particle displacements and velocities from the LPT
initial conditions, relative to their respective standard deviations,
that are very small at the \emph{initial} time.
As before, we vary the floating-point precision, the mesh size, the
particle mass, and the number of time steps.
The order of factor importance and their effects are the same as in the
reproducibility test.
This is because more nonlinear structures are more difficult to reverse.
One way to improve reversibility is use higher order LPT to initialize
the $N$-body simulations at later times \citep{MichauxEtAl2021}, when
the displacements and velocities are not as small.
We leave this for future development.

\begin{table*}[t]
\centering
\caption{\pmwd\ reversibility on GPU.
Our adjoint method reduces memory cost by reconstructing the forward
evolution with reverse time integration.
We test the numerical reversibility by comparing the displacements and
velocities of particles that have evolved to $a=1$ and then reversed to
$a=1/64$, to those of the LPT initial conditions at $a=1/64$, in RMSDs.
We take their ratios to the respective standard deviations, about $0.1
\, \Mpc / h$ and $0.7 \, \mathrm{km} / \mathrm{s}$.
Their smallness is the main reason that the quoted relative differences
here are orders of magnitude greater than those in \autoref{tab:rep}
(see \autoref{fig:grads} where the reversibility is few times more
important than the reproducibility in their impacts on the gradients).
With the same setup as that in \autoref{tab:rep}, we see the same
general trend---left factors are more important than the right ones,
and lower rows are more reversible than the upper ones.
}
\label{tab:rev}
\begin{tabular}{cccccc}
  \toprule
  precision & cell/ptcl & ptcl mass $[10^{10} M_\odot]$ & time steps
    & disp rel diff & vel rel diff \\
  \midrule
  single & 8 & 1 &  63 & $5.2\times10^{-2}$  & $7.1\times10^{-2}$  \\
  single & 8 & 1 & 126 & $2.1\times10^{-2}$  & $3.6\times10^{-2}$  \\
  single & 8 & 8 &  63 & $3.3\times10^{-3}$  & $7.6\times10^{-3}$  \\
  single & 8 & 8 & 126 & $3.7\times10^{-3}$  & $7.0\times10^{-3}$  \\
  single & 1 & 1 &  63 & $1.4\times10^{-3}$  & $2.2\times10^{-3}$  \\
  single & 1 & 1 & 126 & $1.3\times10^{-3}$  & $1.7\times10^{-3}$  \\
  single & 1 & 8 &  63 & $4.3\times10^{-4}$  & $7.3\times10^{-4}$  \\
  single & 1 & 8 & 126 & $4.4\times10^{-4}$  & $6.3\times10^{-4}$  \\
  double & 8 & 1 &  63 & $5.4\times10^{-11}$ & $1.3\times10^{-10}$ \\
  double & 8 & 1 & 126 & $3.5\times10^{-11}$ & $7.0\times10^{-11}$ \\
  double & 8 & 8 &  63 & $5.8\times10^{-12}$ & $1.4\times10^{-11}$ \\
  double & 8 & 8 & 126 & $6.4\times10^{-12}$ & $1.2\times10^{-11}$ \\
  double & 1 & 1 &  63 & $2.2\times10^{-12}$ & $3.4\times10^{-12}$ \\
  double & 1 & 1 & 126 & $2.1\times10^{-12}$ & $2.9\times10^{-12}$ \\
  double & 1 & 8 &  63 & $7.2\times10^{-13}$ & $1.2\times10^{-12}$ \\
  double & 1 & 8 & 126 & $6.9\times10^{-13}$ & $9.4\times10^{-13}$ \\
  \bottomrule
\end{tabular}
\end{table*}

\begin{figure*}[t]
\centering
\hspace{7em}
\includegraphics[width=0.5\linewidth]{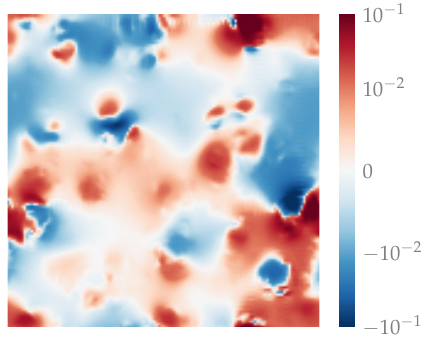}
\\
\vspace{1em}
\includegraphics[height=0.45\linewidth]{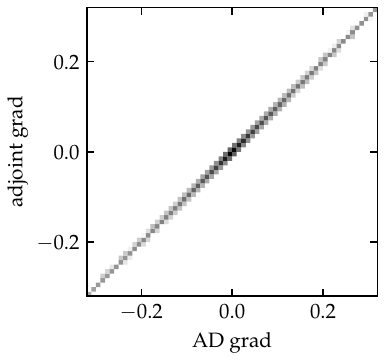}
\hfill
\includegraphics[height=0.45\linewidth]{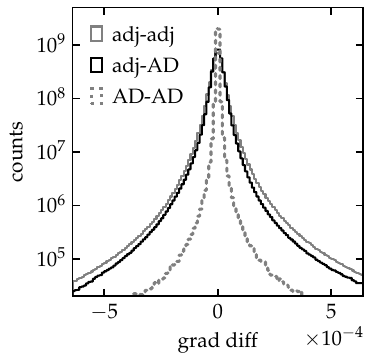}
\caption{Adjoint gradients of a $128\times128$ slice of the real white
noise field $\d\cJ / \d\vomega$ (top panel), in a \pmwd\ simulation of
$128^3$ particles in a $(128\,\Mpc/h)^3$ box, with a $256^3$ mesh, 15
time steps, and single precision.
We choose a mean squared error (MSE) objective between two realizations
with the same cosmology but different initial modes, on their density
fields on the $256^3$ mesh at $a=1$, and then compute the gradients with
respect to one realization while holding the other fixed.
We compare the adjoint gradients to those by AD, for which we have
disabled the custom gradient implementation on the scatter, gather, and
$N$-body time stepping operators.
The adjoint and AD gradients agree as expected, with a RMSD of $\approx
4\times10^{-5}$, 3 orders of magnitude smaller than the standard
deviation of the gradients itself, $\approx 0.015$.
It is also comparable to the difference between two different adjoint
gradients, with a RMSD $\approx 5\times10^{-5}$.
Different AD gradients are more consistent, with a tighter RMSD of
$\approx 1\times10^{-5}$ due to the absence of uncertainty from reverse
time integration.
}
\label{fig:grads}
\end{figure*}

Next, we want to verify that our adjoint method yields the same
gradients as those computed by AD.
As explained in \autoref{sec:adj}, \pmwd\ already utilizes AD on most of
the differentiation tasks.
To get the AD gradients we disable our custom VJP implementations on the
$N$-body time integration, and the scatter and gather operations.
In \autoref{fig:grads}, we compare the adjoint and AD gradients on a
smaller problem, because AD already runs out of memory if we double the
number of time steps or increase the space/mass resolution by
$2^3\times$ from the listed specifications in the caption.
For better statistics, we repeat both adjoint and AD runs for 64 times,
with the same cosmology and white noise modes, and compare their results
by an asymmetric difference of $X_i - Y_j$, where $1 \leq j < i \leq
64$.
First we set $X$ and $Y$ to the adjoint and AD gradients respectively,
and find they agree very well on the real white noise (so do their
gradients on cosmological parameters not shown here; see
\href{https://github.com/eelregit/pmwd/tree/master/docs/papers/adjoint/grads.txt}{\faFile}).
In addition, we can set both $X$ and $Y$ to either adjoint or AD to
check their respective \emph{reproducibility}.
We find both gradients are consistent among different runs of
themselves, with AD being a lot more reproducible without uncertainty
from the reverse time integration but only that from the GPU
reproducibility.
This implies that we can ignore the reproducibility errors
(\autoref{tab:rep}) when the reversibility ones dominate
(\autoref{tab:rev}).
Though this statement should be verified again in the future when we
reduce the reversibility errors using, e.g., 3LPT.
%NOTE: this behavior has disappeared with new JAX/CUDA/GPU
%Interestingly, the asymmetric difference of AD gradients is not
%symmetric about 0, implying some difference between the early and late
%AD results, i.e., possible breaking of order- or time-invariance by
%\texttt{JAX} AD.

\begin{figure*}[t]
\centering
\subfloat[initial conditions at $a=1/64$]{
  \includegraphics[width=0.45\linewidth]{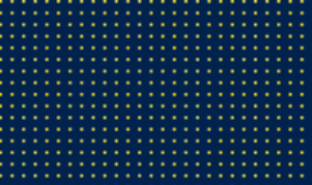}
}
\hfill
\subfloat[final snapshot at $a=1$]{
  \includegraphics[width=0.45\linewidth]{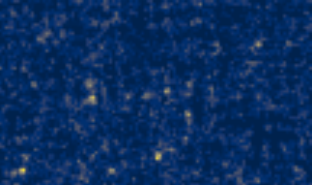}
}
\\
\subfloat[reverse evolution back to $a=1/64$]{
  \includegraphics[width=0.45\linewidth]{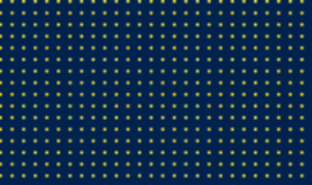}
}
\hfill
\subfloat[optimization for 10 iterations at $a=1$]{
  \includegraphics[width=0.45\linewidth]{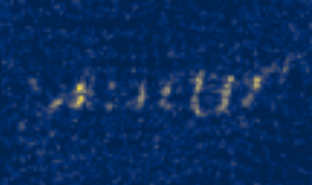}
}
\\
\subfloat[optimization for 100 iterations at $a=1$]{
  \includegraphics[width=0.45\linewidth]{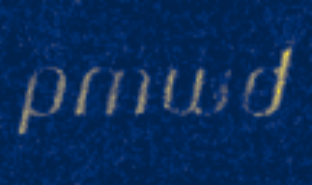}
}
\hfill
\subfloat[optimization for 1000 iterations at $a=1$]{
  \includegraphics[width=0.45\linewidth]{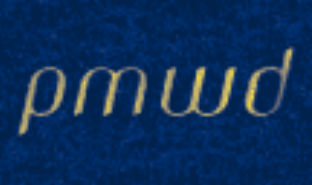}
}
\\
\caption{A toy problem where we optimize the initial conditions by
gradient descent to make some interesting pattern after projection.
The particles originally fill a $16\times27\times16$ grid, and then
evolve from $a=1/64$ to $a=1$ for 63 time steps with \emph{single
precision} and a $32\times54\times32$ mesh in a $160\times270\times160
\, \Mpc^3/h^3$ box.
We compute their projected density in $64\times108$ pixels and compare
that to the target image at the same resolution with an MSE objective.
We use the adjoint method and reverse time integration, assuming the
latter can reconstruct the forward evolution history accurately.
We validate this by demonstrating that the particles evolve backward to
align on the initial grid.
The optimized initial conditions successfully evolve into the target
pattern, which improves with more iterations.
Also see the animated reverse time evolution
\href{https://youtu.be/Epsgh6vr0qs}{\faYoutubePlay} and initial
condition optimization
\href{https://youtu.be/vD6lbjHP3SY}{\faYoutubePlay} on YouTube.
}
\label{fig:optim}
\end{figure*}

Our last test on the adjoint gradients uses them in a toy optimization
problem in \autoref{fig:optim}.
We use the Adam optimizer \citep{Adam} with a learning rate of $0.1$ and
the default values for other hyperparameters.
Holding the cosmological parameters fixed, we can optimize the real
white noise modes to source initial particles to evolve into our target
pattern, in hundreds to thousands of iterations.
Interestingly, we find that the variance of the modes becomes bigger
than 1 (that of the standard normal white noise) after the optimization,
and the optimized modes show some level of spatial correlation not
present in the white noise fields, suggesting that the optimized initial
conditions are probably no longer Gaussian.

\vspace{1em}
\subsection{Performance}

\begin{figure*}[t]
\centering
\includegraphics[width=0.7\linewidth]{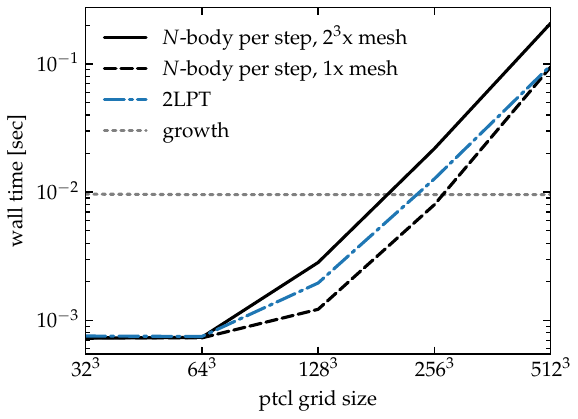}
\caption{Performance of \pmwd. Both the 2LPT and $N$-body components
scale well from $512^3$ particles to $64^3$ particles, below which they
become overhead dominated.
Solving the growth ODEs takes a constant time, and can dominate the cost
for small numbers of particles and few time steps, but generally does
not affect problems with more than $128^3$ particles.
}
\label{fig:perf}
\end{figure*}

\pmwd\ benefits from GPU accelerations and efficient CUDA
implementations of scatter and gather operations.
In \autoref{fig:perf}, we present performance test of \pmwd, and find
both the LPT and the $N$-body parts scale well except for very small
problems.
The growth function solution has a constant cost, and generally does not
affect problem of moderate sizes.
However, for a small number of particles and few time steps, it can
dominate the computation, in which case one can accelerate the growth
computation with an emulator \citep{KwanModiEtAl2022}.

\vspace{1em}
\section{Conclusions}

In this work, we develop the adjoint method for memory efficient
differentiable cosmological simulations, exploiting the reversible
nature of $N$-body Hamiltonian dynamics, and implement it with
\texttt{JAX} in a new PM library, \pmwd.
We have validated the numerical reversibility and the accuracy of the
adjoint gradients.

\pmwd\ is both computation and memory efficient, enabling larger and
more accurate cosmological dark matter simulations.
The next step involves modeling cosmological observables such as
galaxies.
One can achieve this with analytic, semi-analytic, and deep learning
components running based on or in parallel with \pmwd.
In the future, it can also facilitate the simultaneous modeling of
multiple observables and the understanding of the astrophysics at play.

\pmwd\ will benefit all the forward modeling approaches in cosmology,
and will improve gradient based optimization and field-level inference
to simultaneously constrain the cosmological parameters and the initial
conditions of the Universe.
The efficiency of \pmwd\ also makes it a promising route to generate
the large amount of training data needed by the likelihood-free
inference frameworks \citep{CranmerEtAl2020, DELFI}.

Currently, these applications require more development, including
distributed parallel capability on multiple GPUs \citep{FlowPM}, more
accurate time integration beyond \texttt{FastPM} \citep{ListHahn2023},
optimization of spatiotemporal resolution of the PM solvers \citep{LDL,
LanzieriLanusseEtAl2022, ZhangLiEtAl}, short-range supplement by direct
summation of the particle-particle (PP) forces on GPUs \citep{HACC,
PKDGRAV3, ABACUS}, differentiable models for observables, etc.
We plan to pursue these in the future.

\vspace{1em}
\textit{\large Acknowledgements:}
YL is grateful to all organizers of the 2013 KICP Summer School on
Computational Cosmology \citep{KICPsummer2013}, especially Andrey
Kravtsov and Nick Gnedin.
We thank Alex Barnett, Biwei Dai, Boris Leistedt, ChangHoon Hahn,
Daisuke Nagai, Dan Foreman-Mackey, Dan Fortunato, David Duvenaud, David
Spergel, Dhairya Malhotra, Erik Thiede, Francisco Villaescusa-Navarro,
Fruzsina Agocs, Giulio Fabbian, Jens St\"ucker, Kaze Wong, Kazuyuki
Akitsu, Lehman Garrison, Manas Rachh, Ra\'ul Angulo, Shirley Ho, Simon
White, Teppei Okumura, Tomomi Sunayama, Ulrich Steinwandel, Uro\v{s}
Seljak, Wen Yan, Wenda Zhou, William Coulton, Xiangchong Li, and Yueying
Ni for interesting discussions.
YL and YZ were supported by The Major Key Project of PCL.
YZ was supported by the China Postdoctoral Science Foundation under
award number 2023M731831.
The Flatiron Institute is supported by the Simons Foundation.

\vspace{1em}
\textit{\large Code Availability:}
\pmwd\ is open-source on GitHub
\href{https://github.com/eelregit/pmwd}{\faGithub}, including the source
files and scripts of this paper
\href{https://github.com/eelregit/pmwd/tree/master/docs/papers/adjoint}{\faFile}.
Also see the companion code paper \citep{pmwd}
\href{https://github.com/eelregit/pmwd/tree/master/docs/papers/pmwd}{\faFile}.

\vspace{1em}
\textit{\large Software:}
\texttt{JAX} \citep{JAX}, \texttt{NumPy} \citep{NumPy}, \texttt{SciPy}
\citep{SciPy}, and \texttt{matplotlib} \citep{matplotlib}.

\clearpage
\appendix

\vspace{1em}
\section{Growth equations}
\label{app:growth}

The 2LPT growth functions follow the following ODEs:
\begin{align}
D_1'' + \Bigl( 2 + \frac{H'}H \Bigr) D_1' - \frac32 \Omegam(a) D_1
  &= 0, \nonumber\\
D_2'' + \Bigl( 2 + \frac{H'}H \Bigr) D_2' - \frac32 \Omegam(a) D_2
  &= \frac32 \Omegam(a) D_1^2,
\end{align}
where
\begin{equation}
\Omegam(a) = \frac{\Omegam H_0^2}{a^3 H^2}
\end{equation}
If the universe has been sufficiently matter dominated at $a_\ic$, with
$\Omegam(a_\ic) \simeq 1$ and $H'/H \simeq -3/2$, the initial conditions
of the ODEs can be set as the growing mode in this limiting case
\begin{align}
D_1(a_\ic) &= a_\ic, \nonumber\\
D_1'(a_\ic) &= a_\ic, \nonumber\\
D_2(a_\ic) &= \frac37 a_\ic^2, \nonumber\\
D_2'(a_\ic) &= \frac67 a_\ic^2.
\end{align}
$D_1 \simeq a$ in the matter dominated era, before being suppressed by
dark energy.

The above growth equations can be written in such suppression factors,
\begin{equation}
G_m \triangleq D_m / a^m,
\end{equation}
for $m \in \{1, 2\}$, as
\begin{align}
G_1'' + \Bigl( 4 + \frac{H'}H \Bigr) G_1'
  + \Bigl( 3 + \frac{H'}H - \frac32 \Omegam(a) \Bigr) G_1
  &= 0, \nonumber\\
G_2'' + \Bigl( 6 + \frac{H'}H \Bigr) G_2'
  + \Bigl( 8 + 2\frac{H'}H - \frac32 \Omegam(a) \Bigr) G_2
  &= \frac32 \Omegam(a) G_1^2,
\end{align}
with initial conditions
\begin{align}
G_1(a_\ic) &= 1, \nonumber\\
G_1'(a_\ic) &= 0, \nonumber\\
G_2(a_\ic) &= \frac37, \nonumber\\
G_2'(a_\ic) &= 0.
\end{align}
We solve the growth equations in $G_m$ instead of $D_m$, with the
\texttt{JAX} adaptive ODE integrator implementing the adjoint method in
the optimize-then-discretize approach.
This is because the former can be integrated backward in time more
accurately for early times, which can improve the adjoint gradients.

We can then evaluate the \texttt{FastPM} time integration factors in
\eqref{G_ZA} by
\begin{align}
G_D &= a^2 \dot D_1 = a^2 H D_1', \nonumber\\
G_K &= a \dot G_D = a^3 H^2 \Bigl[
  D_1'' + \Bigl(2 + \frac{H'}H \Bigr) D_1' \Bigr],
\label{G_eval}
\end{align}
and
\begin{align}
D_m &= a^m G_m \nonumber\\
D_m' &= a^m \bigl( m G_m + G_1' \Bigr), \nonumber\\
D_m'' &= a^m \bigl( m^2 G_1 + 2m G_1' + G_1'' \Bigr).
\end{align}

\bibliographystyle{aasjournal}
\bibliography{adjoint}

%\listofchanges

\end{document}